\documentclass[conference,compsoc]{IEEEtran}
\ifCLASSOPTIONcompsoc
  \usepackage[nocompress]{cite}
\else
  \usepackage{cite}
\fi
\usepackage{hyperref}        

\ifCLASSINFOpdf
  \usepackage[pdftex]{graphicx}
\else
\fi
\usepackage{amssymb}

\usepackage{amsmath}

\usepackage{algorithmic}
\usepackage[ruled,vlined]{algorithm2e} 

\usepackage{booktabs}

\begin{document}
%
\title{RL-Finetuned LLMs for Privacy-Preserving Synthetic Rewriting}

\author{
    Zhan Shi, Yefeng Yuan, Yuhong Liu, Liang Cheng, Yi Fang \\
    \textit{Santa Clara University, eBay} \\
    \{ashi2, yyuan4, yhliu, yfang\}@scu.edu, liacheng@ebay.com
}

\maketitle

\begin{abstract}

The performance of modern machine learning systems depends on access to large, high-quality datasets, often sourced from user-generated content or proprietary, domain-specific corpora. However, these rich datasets inherently contain sensitive personal information, raising significant concerns about privacy, data security, and compliance with regulatory frameworks. While conventional anonymization techniques can remove explicit identifiers, such removal may result in performance drop in downstream machine learning tasks. More importantly, simple anonymization may not be effective against inference attacks that exploit implicit signals such as writing style, topical focus, or demographic cues, highlighting the need for more robust privacy safeguards during model training. To address the challenging issue of balancing user privacy and data utility, we propose a reinforcement learning framework that fine-tunes a large language model (LLM) using a composite reward function that jointly optimizes for explicit and implicit privacy, semantic fidelity, and output diversity. To effectively capture population level regularities, the privacy reward combines semantic cues with structural patterns derived from a minimum spanning tree (MST) over latent representations. By modeling these privacy-sensitive signals in their distributional context, the proposed approach guides the model to generate synthetic rewrites that preserve utility while mitigating privacy risks. Empirical results show that the proposed method significantly enhances author obfuscation and privacy metrics without degrading semantic quality, providing a scalable and model-agnostic solution for privacy preserving data generation in the era of large language models.
\end{abstract}


%
\IEEEpeerreviewmaketitle

\section{Introduction}
The advent and rapid proliferation of Artificial Intelligence (AI) have underscored the critical role of data in training effective machine learning models. These sophisticated models, capable of human-like text generation and comprehension, are fueled by vast datasets, often encompassing personal or domain-specific information to meet the increasing requirements of downstream applications and specialized tasks. This insatiable demand for high-quality data, however, runs parallel to mounting concerns regarding user privacy, as recent research demonstrates that models can inadvertently leak sensitive information, even when direct identifiers are masked \cite{meng2025rr}. Particularly, there are two major challenges.

First, \textbf{implicit privacy} \cite{yang2022implicit} can be inferred from seemingly innocuous text by advanced attacks. Unlike explicit personally identifiable information (PII), such as names or addresses, implicit attributes (e.g. gender, age, religion, personal interests, etc.) may be deduced from linguistic nuance. For instance, the sentence “\textit{Just grabbed a late-night bite at that new taco truck after coding all day}” contains no direct location marker, yet a motivated attacker could infer the author’s likely residence in a tech-centric city like San Jose based on cultural context and lifestyle cues. 

Second, \textbf{stylometric leakage} magnifies this risk. Even after explicit PII is scrubbed, an author’s \emph{writing fingerprint} (i.e., features like function-word frequency, sentence-length distribution, punctuation preferences, and higher-order discourse patterns), remains remarkably stable \cite{yukhymenko2024synthetic}. Stylometry models routinely achieve $>90\%$ accuracy in re-identifying authors given only a handful of paragraphs. Moreover, extreme stylistic or lexical anomalies (statistical outliers) attract adversarial attention: they sit far from population manifolds, making them high-value targets for attribute inference attacks.

Existing industry-standard anonymization tools, which primarily focus on pattern-matching for structured PII,
often fall short in addressing complex, context-dependent implicit and stylometric privacy leakages \cite{lukas2023analyzing}. Furthermore, these anonymization tools often indiscriminately mask information, leading to a loss of data utility and readability. It is particularly problematic for the short, informal nature of much online text. On the other hand, Differential Privacy (DP) \cite{abadi2016deep} offers formal guarantees, but lowering the privacy budget $\epsilon$ often degrades semantic quality, forcing practitioners into a harsh privacy–utility trade-off. These challenges coupled with the increasing regulatory push for stronger privacy protections\cite{chim-etal-2025-evaluating} (e.g., GDPR, CCPA), are driving demand for new robust user privacy protection strategies. 

Recent advancement in generative AI highlights a promising direction: protect user privacy by replacing the original sensitive contents with text-rewriting \cite{meisenbacher2024justrewriteagainpostprocessing}. For example, some studies propose adversarial approaches, where one LLM attempts to anonymize text while another tries to infer attributes\cite{staab2023beyond}, to enhance the robustness of synthetic text against privacy inference attacks. However, state-of-the-art privacy-preserving text generation methods typically rewrite a \emph{single} input under a global semantic-consistency objective. Consequently, (1) global stylistic cues persist, (2) outliers remain unsmoothed, and (3) the lack of diverse candidates precludes selecting the least recognizable version.

\textbf{Threat Model} In this work, we mainly focus on privacy attacks where malicious attackers aim to infer users' sensitive information, including not only PII, but also other attributes, such as age, gender, location, personal interests, online writing style, etc. By collecting such information, malicious attackers may be able to perform direct attacks, such as identity theft, impersonation, and singling out attacks, or indirect misuse such as target marketing or sales of personal data. 

To address such privacy threats, we propose a \textsc{STYLE-MST Rewriter}, an LLM-based privacy-aware text-rewriting framework that integrates corpus-level graph regularization with preference-based reinforcement learning.  The proposed framework maintains, in streaming fashion, a minimum–spanning tree (MST), whose edge weights depend jointly on TF–IDF \(n\)-gram vectors (capturing surface semantics) and learned style embeddings (capturing authorial voice). A candidate rewrite is rewarded when its average \emph{NovelSum} \cite{yang2025measuring}, a density-aware measure of local semantic novelty, is high and its contribution to the tree’s mean edge length is low, thereby discouraging both mode collapse and isolated stylistic outliers.  A memory-aware instance of Group Relative Policy Optimization (GRPO) \cite{shao2024deepseekmath}trains a Low-Rank Adaptation (LoRA) \cite{hu2022lora} adapter attached to a frozen 3-B-parameter \textsc{LLaMA} \cite{touvron2023llama} backbone.  For every prompt the model emits four rewrites together with an explicit reasoning trace that lists detected PII and planned transformations; the highest-valued rewrite and a randomly selected lower-valued rewrite form a preference pair that drives the GRPO update.  The composite reward combines BERT-based semantic similarity, explicit-entity overlap, style-entropy derived from a lightweight AuthorMix classifier, length and format regularisers, and the graph-based STYLE–MST term. The major contributions of this work are summarized below.



\begin{itemize}
    \item First, we craft a graph-regularized \textsc{STYLE-MST} reward that promotes semantic diversity while suppressing stylistic outliers through population-level regularization that explicitly incorporates contextual information. 
    \item Second, we inject chain-of-thought reasoning into a reinforcement-learning procedure that optimizes over multiple candidate rewrites, enabling amortized inference and superior generalization beyond greedy reward selection.
    \item Third, we pair an off-the-shelf LLM with a multi-axis scoring module that jointly evaluates semantic fidelity, explicit PII exposure, implicit attribute leakage, stylometric similarity, and outlier risk, returning the safest yet most faithful paraphrase. 
\end{itemize}

This unified reward schema balances five privacy dimensions within a lightweight, high-throughput pipeline, achieving a significant reduction in author reidentification accuracy while preserving the original semantics compared with strong baselines. 
Together, these components deliver high-utility controllable rewriting that retains meaning and stylistic coherence across diverse tasks, closing the introduction to our study.

The rest of the paper is organized as follows. Related work is discussed in Section \ref{sec:relatedwork}, followed by an overview of the proposed framework in Section \ref{sec:framework}. Then the three major components of the proposed framework are discussed in Section \ref{sec:prompt_construct} - \ref{sec:RL_trainingloop}. The experiment setup and validation of the proposed scheme is discussed in detail in Section \ref{sec:exp} - \ref{sec:result_analysis}, and Section \ref{sec:conclusion} concludes the paper. 

\begin{figure*}[!t] 
    \centering
    \includegraphics[width=\textwidth,keepaspectratio]{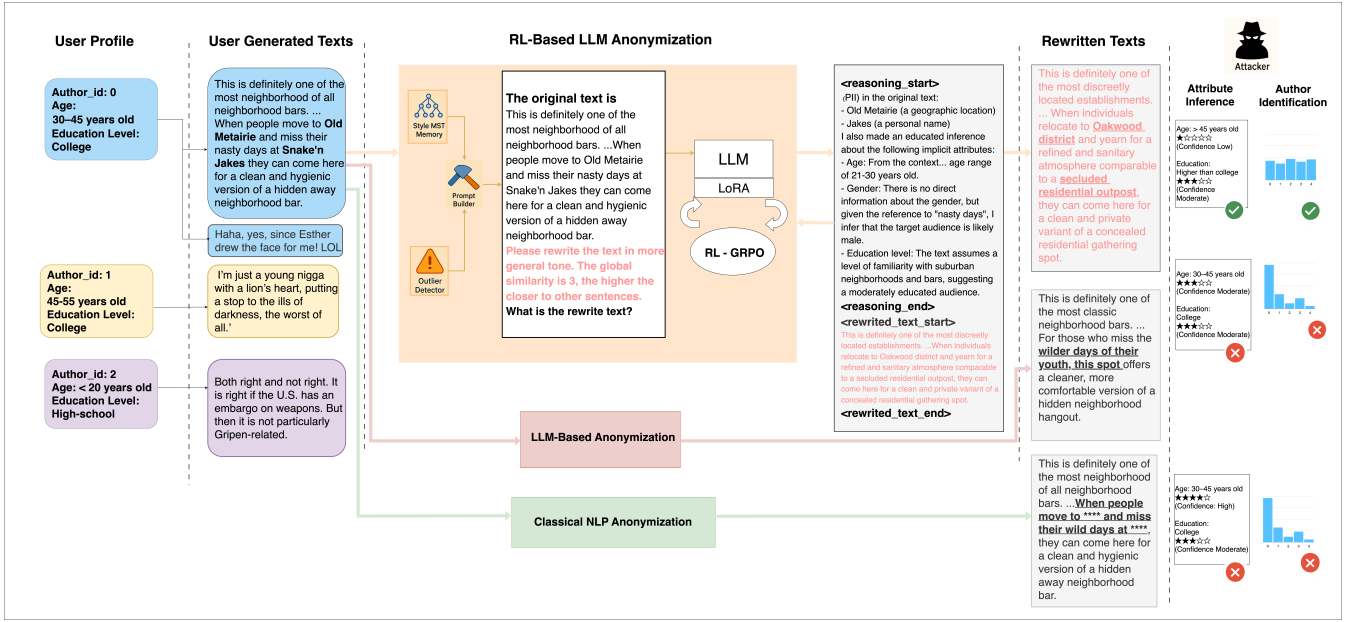}
    \caption{Overview of RL-Based LLM Anonymization Framework.The system takes user-generated texts tied to user profiles (e.g., author ID, age, education level) and rewrites them using a reinforcement learning (RL)-enhanced LLM anonymizer. The core module integrates signals from a Style-MST memory and an outlier detector to construct prompts that encourage generalization. The reward function balances privacy (via PII removal and style obfuscation), semantic fidelity, and formatting consistency. Compared to classical anonymization (e.g., regex or NER-based), our approach produces rewrites that are semantically coherent while minimizing author re-identification and attribute inference risks, as illustrated in the rightmost attacker evaluation.}
    \label{fig:top_descripe_image}
\end{figure*}

\section{Related Work}
\label{sec:relatedwork}
\subsection{Authorship Obfuscation (AO)}
Early AO methods employed rule-based transformations, such as synonym substitution and contraction expansion~\cite{potthast-2016-author,karadjov2017caseaveragemediocrityapproach}. While computationally efficient, these approaches often degraded grammatical fluency and semantic coherence.

Subsequent research framed AO as an adversarial attack on authorship attribution (AA) or verification (AV) systems. Techniques like HotFlip~\cite{ebrahimi2018hotflipwhiteboxadversarialexamples} and black-box perturbation~\cite{gao2018blackboxgenerationadversarialtext} sought to minimally alter inputs to mislead classifiers while preserving semantics. However, such perturbations frequently impair readability~\cite{Crothers_2022}.

Back-translation-based rewriting has also been explored~\cite{altakrori-etal-2022-multifaceted,deyoung2020eraserbenchmarkevaluaterationalized}, offering effective style alteration at the cost of potential semantic drift and unnatural phrasing.

Deep generative models such as VAEs and GANs have been used to disentangle content and style for obfuscation~\cite{shetty2018a4ntauthorattributeanonymity,mireshghallah2021privacyregularizationjointprivacyutility}. More targeted approaches like Alison~\cite{xing2024alisonfasteffectivestylometric} manipulate POS and character n-gram patterns to confuse stylometric models.

StyleRemix~\cite{fisher-etal-2024-styleremix} introduced an interpretable LoRA-based framework that adjusts writing across axes such as formality, sentiment, and narrativity. However, such population-level manipulations may not fully mitigate individual stylistic signatures.

\subsection{Privacy-Preserving Text Generation}
LLMs have heightened privacy risks. Carlini et al.~\cite{carlini2021extractingtrainingdatalarge} demonstrated that GPT-2 can memorize and leak training data. Beyond memorization, inference attacks can extract sensitive attributes from model outputs~\cite{staab2024memorizationviolatingprivacyinference}.

Differential Privacy (DP) has been employed to mitigate these risks. Central DP methods like DP-SGD~\cite{abadi2016deep} perturb gradients during training to provide privacy guarantees but often degrade utility. Local DP methods inject noise at the token or embedding level~\cite{fernandes2019generalised,vu2024granularitycrucialapplyingdifferential}.

DP-MLM~\cite{meisenbacher2024dpmlmdifferentiallyprivatetext} proposes a differentially private masked language model trained with formal guarantees, offering stronger privacy during generation. DIPPER~\cite{krishna2024paraphrasing} introduces a controllable paraphrasing framework that disentangles utility and privacy dimensions, allowing for tunable rewriting.

Prompt sanitization approaches, such as Casper~\cite{chong2024casper}, use regex patterns, named entity recognition, and local models to suppress PII before inference. 

Recent efforts generate privatized outputs directly from LLMs. Mattern et al.~\cite{mattern2022differentially} introduced DP-noised GPT-2 outputs, while Igamberdiev and Habernal~\cite{igamberdiev2023dp} extended this to BART. DP-Prompt~\cite{utpala2023locally} achieves privatization via prompt randomization without model retraining. However, these methods often lack token-level budget control and sacrifice generation fidelity.

Tripto et al.~\cite{meisenbacher2024justrewriteagainpostprocessing} observed that repeated paraphrasing with LLMs induces stylistic convergence while largely preserving semantics, motivating approaches that incorporate memory-aware diversity mechanisms.

\section{Framework Overview} \label{sec:framework}
\begin{figure}
    \centering
    \includegraphics[width=1\linewidth]{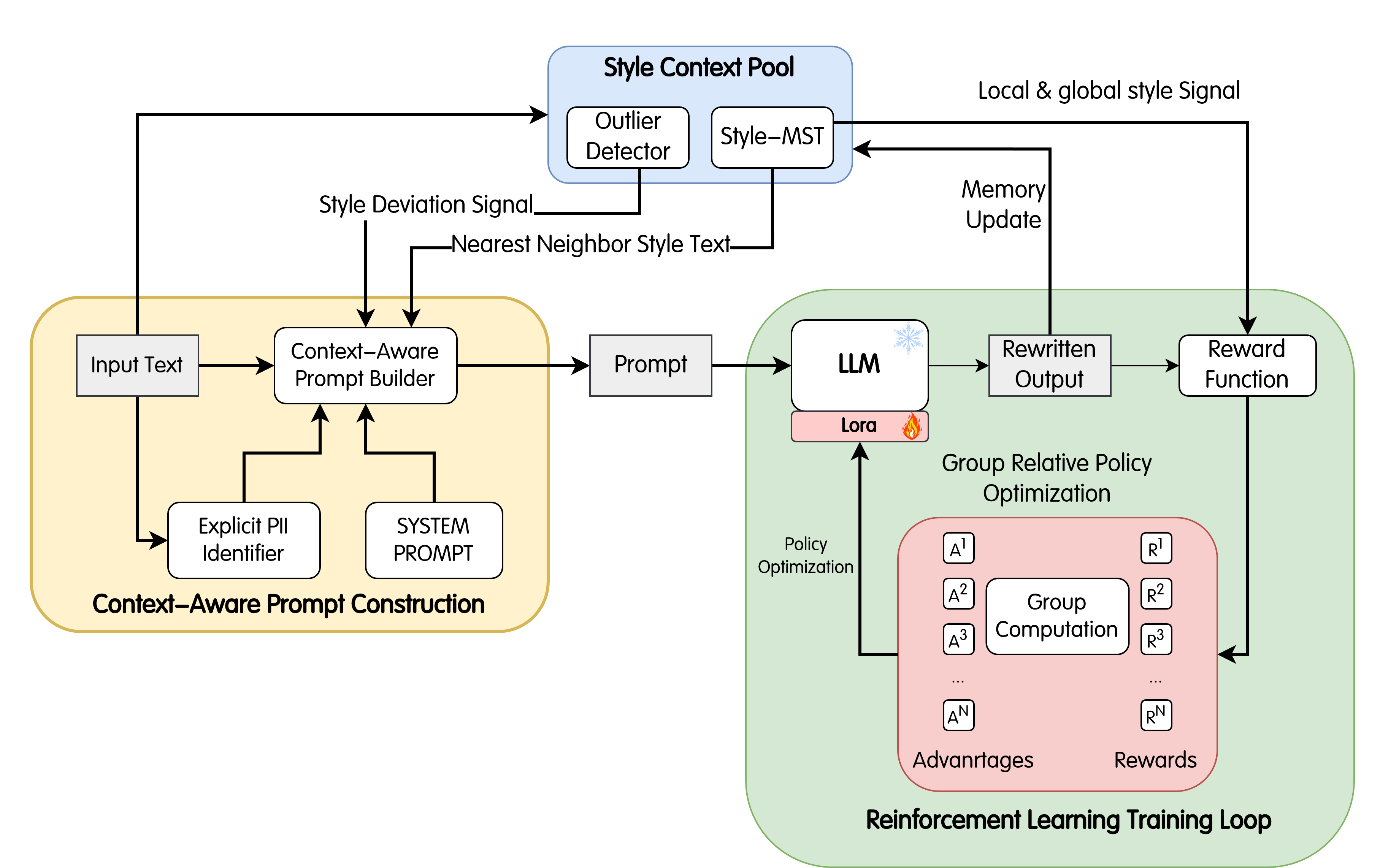}
    \caption{Architecture of the RL-Based Style-Aware Anonymization Framework.
The pipeline consists of a preprocessing module, a global style memory, and a reinforcement learning (RL) training loop.}
    \label{fig:rl_diagram}
\end{figure}



\noindent
In this work, we propose a privacy-preserving text rewriting framework that leverages reinforcement learning to guide large language models (LLMs) in generating synthetic, privacy-compliant outputs. Figure~\ref{fig:rl_diagram} illustrates the end-to-end architecture of our framework, which is structured into three major components: 

\textbf{Context-Aware Prompt Construction}
For each input text, we perform context-aware prompt construction to extract features essential for privacy preserving rewriting. This involves identifying explicit PII and evaluating the stylistic properties of the text with respect to previously generated samples. Specifically, we determine whether the input is a stylistic outlier and retrieve its most similar sentence in terms of style from memory. These insights are used to construct a comprehensive prompt composed of: (1) a system-level instruction template, (2) contextual signals derived from the input and style memory, and (3) the original user text. The final prompt is then submitted to the LLM for rewriting.

\textbf{Style Context Pool}
The \emph{Style Context Pool} maintains a dynamic repository of previously seen inputs and model outputs, serving as a reference for stylistic grounding and diversity promotion. It supports two key modules: (1) a \textit{Style Outlier Detector} that identifies atypical stylistic patterns using distance-based novelty scoring, and (2) a \textit{STYLE-MST Module}, which constructs a minimum spanning tree over the style embedding space of historical generations to model global style relationships and promote stylistic coverage \cite{yuan2024multifacetedevaluationframeworkassessing}. Signals derived from both modules inform the prompt construction process, helping modulate tone, generality, and abstraction during rewriting.

\textbf{Reinforcement Learning Training Loop}
The constructed prompt is provided to a LoRA-augmented \cite{hu2022lora} LLM, which generates multiple rewrite candidates for each input. Each candidate includes both a reasoning trace and a rewritten version of the original text. A composite reward function is then applied to evaluate the candidates along multiple dimensions: semantic consistency, brevity, explicit preservation of privacy, and Style-MST-based stylistic diversity (for implicit protection of privacy). These reward signals are used in Group Relative Policy Optimization (GRPO) \cite{shao2024deepseekmath} to update LoRA adapter weights. The highest-scoring outputs are retained and used to update both the Outlier Detector and the STYLE-MST modules within the Style-Aware Context Pool, closing the RL feedback loop and enhancing context for future generations.

\section{Context-Aware Prompt Construction}\label{sec:prompt_construct}

\subsection{Context Signal Extraction}
\label{PrivacyContextSignalExtraction}
To guide the rewriting process with personalized privacy controls and stylistic alignment, we extract a rich context set $C(x)$ for each input text $x$.

\textbf{Explicit PII Identification:}  
We first identify surface-level PII, including entities such as names, locations, dates, and contact details, using an entity recognition module. The extracted entities are formatted as a structured list and embedded into the prompt to explicitly instruct the model to obfuscate or generalize these attributes during rewriting.

\textbf{Style Deviation Signal:}  
To account for implicit privacy risks, we evaluate the stylistic deviation of the input relative to previous samples using the \texttt{Outlier Detector} module. This module determines whether the input is a stylistic outlier. If flagged, the prompt includes a directive that encourages a more neutral or generic tone. Otherwise, the model is allowed to retain stylistic diversity. This mechanism reduces the likelihood of author re-identification while promoting style-aware generalization.

\textbf{Nearest Neighbor Style Text:}  
To provide global stylistic grounding, we retrieve the closest stylistic match from the \texttt{Style-MST}. The nearest neighbor's rewriting is fed into the prompt directly as context learning,  offering in-distribution stylistic guidance. This context anchor helps mitigate stylistic drift while reinforcing population-level regularity, enhancing both safety and coherence in the rewritten output.

\subsection{Prompt Construction}
To guide the LLM in generating stylistically diverse and privacy-preserving rewrites, we design the prompt using three modular components. Each component serves a distinct role in shaping the model's behavior, ensuring accurate content transformation with privacy and stylistic alignment in mind.

\textbf{System Prompt}  
The system prompt defines the global behavior of the LLM, defining it as a \emph{privacy-preserving rewriting assistant}. It outlines task-level expectations such as removing personal identifiers, adhering to a specified tone (e.g., neutral or professional), and maintaining a consistent output format. In addition, it may instruct the model to include structured reasoning traces or to clearly delimit output segments. A full example of the system prompt is provided in Appendix~\ref{appendix:system_prompt}.

\textbf{Context Prompt}  
The context prompt incorporates contextual signals $C(x)$ extracted from the input, as described in Section~\ref{PrivacyContextSignalExtraction}. These signals include whether the input is identified as a stylistic outlier and the nearest stylistic neighbor retrieved from the memory pool. This context informs the model about whether to neutralize the writing style or maintain diversity by aligning with in-distribution stylistic patterns. This module connects general rewriting rules from the system prompt with instance-specific privacy cues, allowing for fine-grained control over the rewriting process.

\textbf{Final Prompt}  
The final component presents the user-generated content to be rewritten. This input may contain explicit or implicit privacy-sensitive information. Depending on the preprocessing step, entities may be presented in raw form or replaced with placeholders. When combined with the system and context prompts, the complete instruction provided to the model is:

\begin{quote}
\texttt{\{system\_prompt\} The original text is: \{x\} \textbackslash n Detected PII: \{pii\_infos\} \textbackslash n The text is a stylistic outlier, with closest writing context: \{style\_context\} \textbackslash n What is the rewritten text?}
\end{quote}

\section{Style Context Pool} \label{sec:style_context}
\subsection{Stylistic Embeddings and Author Correlation}
\label{sec:privacy_utility}

Writing style carries rich implicit signals that can strongly correlate with personal attributes such as age, gender, and identity. In privacy-preserving text rewriting, a key challenge is to obscure such authorial signals while preserving the original meaning of the content. This requires a careful trade-off between two competing goals: protecting user privacy and maintaining semantic utility.

To address this, our framework decomposes the latent representation space into two orthogonal components: one capturing semantic meaning and the other modeling stylistic traits. We leverage this decomposition to guide the generation of rewritten text that is faithful in meaning but obfuscated in style.

\textbf{Latent Subspace Decomposition.}
We define the representation space of a text sample as composed of:
\begin{itemize}
    \item \textbf{Semantic Subspace:} This subspace captures the core informational content of the text. We obtain semantic embeddings using contextual encoders such as BERT, denoted by $f_{\text{sem}}(\cdot)$, which are known to encode meaning while being relatively agnostic to surface-level style.
    
    \item \textbf{Stylistic Subspace:} This subspace encodes authorial traits—such as tone, lexical choice, and syntactic preferences—that often carry re-identifiable signals. We model this space using our proposed \textbf{STYLE-MST}, denoted by $f_{\text{style}}(\cdot)$, which is constructed to enhance sensitivity to author-specific stylistic patterns while promoting structural diversity.
\end{itemize}

\begin{figure}
    \centering
    \includegraphics[width=1\linewidth]{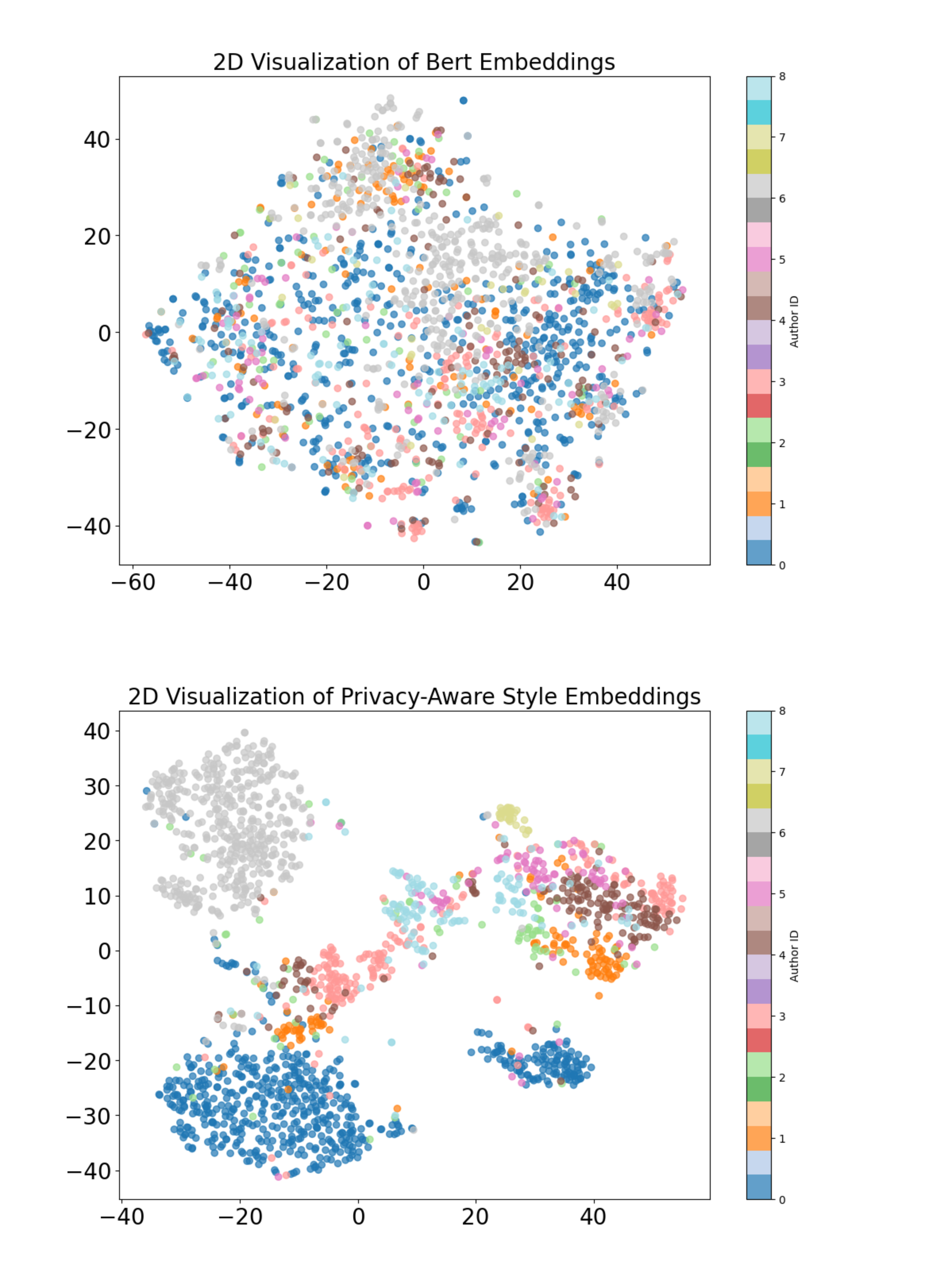}
    \caption{
    \textbf{t-SNE Visualization of Embedding Spaces Colored by Author ID.} 
    Top: BERT-based embeddings show weak clustering and minimal author separation, reflecting their focus on semantic content. 
    Bottom: Our privacy-aware style embeddings, trained with author identification supervision, reveal clear and discriminative clusters aligned with individual authors. This demonstrates the effectiveness of the proposed STYLE-MST embedding in capturing authorial style and its potential for stylometric privacy risk modeling.
    }
    \label{fig:style_embedding}
\end{figure}

\textbf{Optimization Objective.}
Our goal is to transform an input text $x$ into a rewritten version $y$ that satisfies the following:
\begin{enumerate}
    \item \textbf{High Semantic Fidelity:} $y$ should remain close to $x$ in the semantic space, ensuring preservation of the original content.
    \item \textbf{High Stylistic Divergence:} $y$ should move away from $x$ in the stylistic space to mask identity and reduce stylometric leakage.
\end{enumerate}

To formalize this trade-off, we define a privacy-utility loss that penalizes semantic drift while encouraging stylistic deviation:
\begin{equation}
\mathcal{L}(x, y) = \underbrace{\left\|f_{\text{sem}}(x) - f_{\text{sem}}(y)\right\|_2}_{\text{Utility Loss}} - \gamma \, \underbrace{\left\|f_{\text{style}}(x) - f_{\text{style}}(y)\right\|_2}_{\text{Privacy Gain}}
\label{eq:privacy_utility_loss}
\end{equation}
Here, $\gamma$ is a tunable hyperparameter that controls the strength of privacy enforcement relative to utility preservation.

\textbf{Learning Privacy-Aware Style Embeddings.}
While pre-trained models like \cite{patel2025styledistancestrongercontentindependentstyle} and \cite{wegmann-etal-2022-author} offer good style representations, they often entangle meaning not strongly correlated with author identification. To obtain more disentangled and privacy-sensitive stylistic embeddings, we train a Multi-Layer Perceptron (MLP) over the output of a frozen style encoder using author classification as a supervisory signal. Specifically, we fine-tune the stylistic projection on the Blog Authorship Corpus, using author IDs as labels. This approach yields embeddings that are highly discriminative with respect to authorial identity, making them well-suited for guiding style-based anonymization and evaluating stylistic divergence.

\textbf{Embedding Visualization.}
Figure~\ref{fig:style_embedding} compares the 2D t-SNE projections of standard BERT embeddings (left) with our proposed privacy-aware style embeddings (right), colored by author ID. The BERT-based \cite{liu2019roberta} embeddings show minimal separation between authors, indicating weak author discrimination. In contrast, our style embedding space exhibits clear author-specific clusters, revealing strong alignment between stylistic representations and author identity. This illustrates both the need for style obfuscation in privacy-preserving rewriting and the effectiveness of our style embedding in modeling these authorial signals.

By embedding the privacy-utility trade-off directly into the generation policy itself, our method learns to produce high-utility outputs that are resilient to both explicit and implicit privacy threats.

\subsection{Stylistic Outlier Detection}
\label{sec:outlier_detection}

While explicit identifiers are often the primary focus of privacy preservation, implicit signals, particularly an author's unique writing style can expose them to sophisticated stylometric re-identification attacks. To counter this threat, we introduce a principled mechanism for detecting stylistically atypical inputs that carry a higher risk of identity leakage.

We represent each input text $x$ as a high-dimensional vector $\mathbf{v}_x \in \mathbb{R}^d$ using a SentenceTransformer model specifically fine-tuned to capture stylistic features like tone, syntax, and expression. Given a reference corpus of prior texts $\mathcal{X} = \{x_1, \dots, x_n\}$ with corresponding style embeddings $\mathbf{V} = \{\mathbf{v}_1, \dots, \mathbf{v}_n\}$, we quantify the stylistic uniqueness of a new input $x$ by computing its average cosine distance to the corpus:
\begin{equation}
\bar{d}_x = \frac{1}{n} \sum_{i=1}^{n} \left( 1 - \cos(\mathbf{v}_x, \mathbf{v}_i) \right),
\end{equation}
where $\cos(\cdot, \cdot)$ denotes the cosine similarity. This metric measures how much an input's style deviates from the established norm of the corpus. Inputs that lie far from the centroid of this style distribution are considered potential outliers.

To formalize this, we define an adaptive outlier threshold $\tau$ based on the empirical distribution of these distances:
\begin{equation}
\tau = \mu + \lambda \cdot \sigma,
\end{equation}
where $\mu$ and $\sigma$ are the mean and standard deviation of the average distances $\{\bar{d}_i\}_{i=1}^n$ across the reference corpus $\mathcal{X}$, and $\lambda$ is a tunable hyperparameter (default: $\lambda=2.0$) that controls the sensitivity of the detector.

If $\bar{d}_x > \tau$, the input is flagged as a \textbf{stylistic outlier}. Such texts are presumed to contain distinctive and user-specific stylistic signals. When an outlier is detected, our system adaptively modifies the generation policy by conditioning the LLM on a rewriting prompt that emphasizes generality, neutrality, and a formal tone, thereby discouraging the model from replicating the unique and potentially identifying stylistic traits. For in-distribution texts ($\bar{d}_x \le \tau$), the system preserves stylistic diversity to promote linguistic richness and naturalness.

\begin{figure}[!h]
    \centering
    \includegraphics[width=0.8\linewidth]{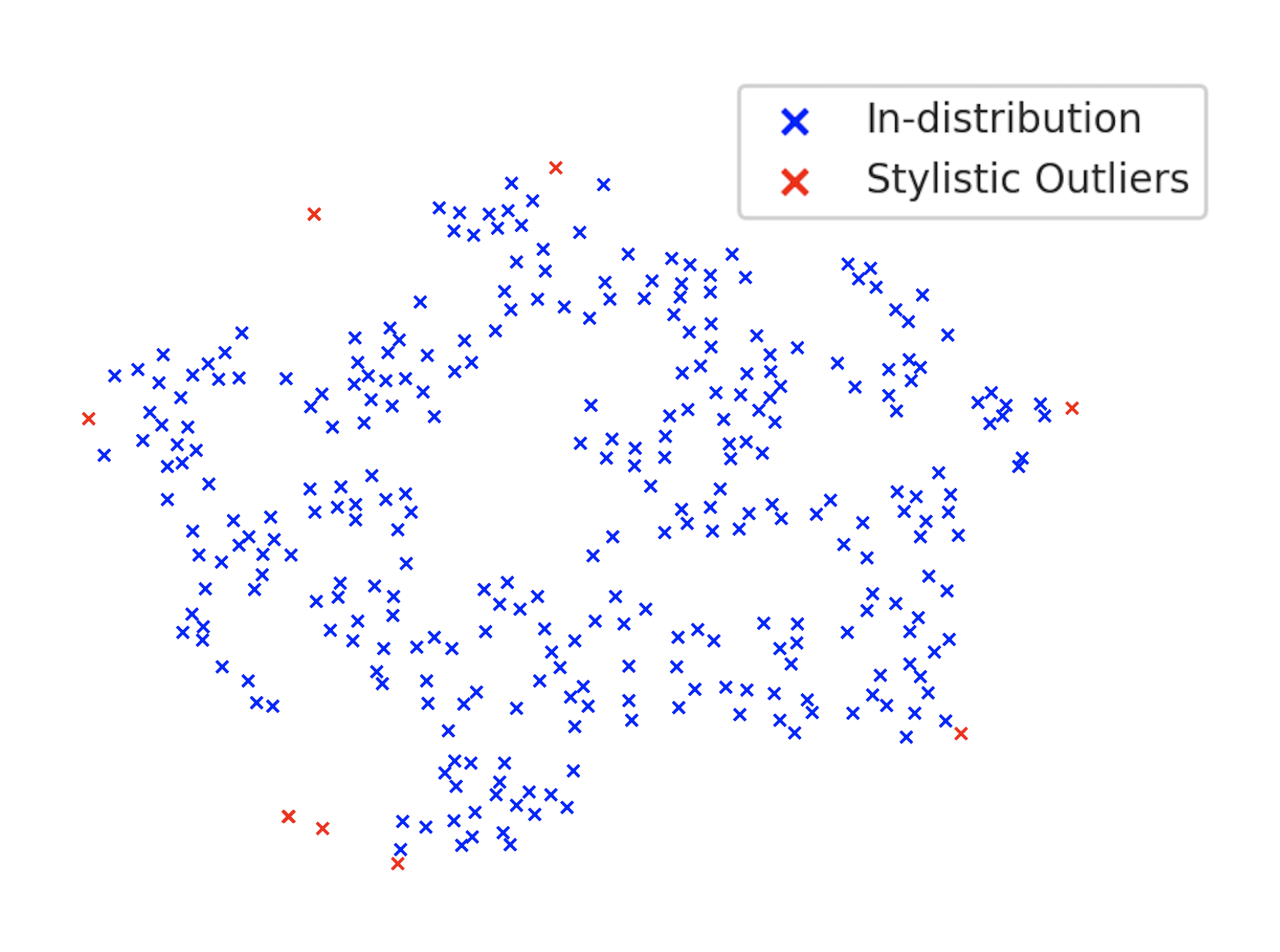}
    \caption{2D t-SNE visualization of style embeddings. Stylistic outliers (red) are clearly separated from the dense main corpus cluster (blue), reflecting distinct authorial traits that may pose implicit privacy risks.}
    \label{fig:style_outlier_tsne}
\end{figure}


\subsection{STYLE-MST: Graph-Regularized Diversity and Privacy Control}
\label{sec:style_mst}

To manage stylistic diversity and enforce implicit privacy constraints at a population level, we introduce STYLE-MST, a novel graph-based reward mechanism. Unlike traditional token-level rewards (e.g., BLEU), STYLE-MST imposes a structural constraint on the entire set of generated outputs. It does this by organizing previously generated sentences into a dynamic Minimum Spanning Tree (MST), denoted $\mathcal{T}$, constructed within the style embedding space.

\begin{algorithm}[t]
\caption{Distance-Based Stylistic Outlier Detection}
\label{alg:style-outlier}
\KwIn{Input texts $\mathcal{X} = \{x_1, x_2, \dots, x_n\}$, embedding function $\texttt{embed}$, radius threshold $r$, minimum neighbor count $k$, Z-score threshold $\lambda$}
\KwOut{Outlier flags $\mathbf{o} \in \{0,1\}^n$, average distance scores $\{\bar{d}_i\}_{i=1}^n$}

\vspace{0.5em}
$\mathbf{V} \leftarrow \texttt{embed}(\mathcal{X})$ 
\For{$i \gets 1$ \KwTo $n$}{
    $\texttt{distances} \leftarrow \{1 - \cos(\mathbf{v}_i, \mathbf{v}_j) \mid j \neq i\}$ 
    $\mathcal{N}_i \leftarrow \{j \mid \texttt{distances}[j] < r\}$ 
    \uIf{$|\mathcal{N}_i| < k$}{
        $o_i \leftarrow 1$ 
    }
    \Else{
        $\bar{d}_i \leftarrow \frac{1}{|\mathcal{N}_i|} \sum_{j \in \mathcal{N}_i} \texttt{distances}[j]$ 
        $o_i \leftarrow 0$ 
    }
}
$\mu \leftarrow \text{mean}(\{\bar{d}_i\}_{i \mid o_i = 0})$ 
$\sigma \leftarrow \text{std}(\{\bar{d}_i\}_{i \mid o_i = 0})$ 
$\tau \leftarrow \mu + \lambda \cdot \sigma$ 

\For{$i \gets 1$ \KwTo $n$}{
    \If{$\bar{d}_i > \tau$}{
        $o_i \leftarrow 1$ 
    }
}
\Return $\mathbf{o}, \{\bar{d}_i\}_{i=1}^n$
\end{algorithm}

\subsubsection{Dynamic Stylistic Tree Construction}

Each node in the MST $\mathcal{T}$ corresponds to the style embedding $\mathbf{v}_i$ of a previously accepted output. The edge weight between any two nodes $i$ and $j$ is defined by their stylistic dissimilarity:
\begin{equation}
w_{ij} = 1 - \mathrm{cos}(\mathbf{v}_i, \mathbf{v}_j),
\end{equation}
where $\mathrm{cos}(\cdot, \cdot)$ is the cosine similarity in the style space. As new candidate outputs are generated, the tree is dynamically updated according to Algorithm~\ref{alg:stylemst}.

This algorithm determines whether a new sentence $s$ is stylistically novel enough to be inserted as a new branch, thereby enriching diversity, or if it is similar enough to an existing node to be merged, reinforcing stylistic conformity and reducing author traceability.

\subsubsection{Reward Computation and RL Integration}

The STYLE-MST structure is used to compute a reward signal, $r_5(y)$, which has two key adaptive components:
\begin{itemize}
    \item \textbf{Global Deviation}: This component measures the average similarity score between the candidate $y$ and all nodes in the tree $\mathcal{T}$. It encourages exploration by penalizing rewrites that are too similar to past outputs.
    \item \textbf{Local Density Adjustment}: This component evaluates similarity within a local neighborhood of the candidate's style embedding. It provides a fine-grained signal about whether the candidate is moving into a dense (stylistically common) or sparse (stylistically unique) region of the space.
\end{itemize}
Crucially, this reward is modulated by the \textbf{stylistic outlier detector} (Section~\ref{sec:outlier_detection}). If the original input $x$ was flagged as an outlier, the reward function incentivizes the rewrite $y$ to move towards high-density regions of the MST, effectively neutralizing its stylistic signature. If $x$ is in-distribution, the reward favors stylistic spread, encouraging expressive and diverse outputs.

This graph-regularized reward is integrated as a key component into the composite reward function for our RL agent:
\begin{equation}
R(y) = \sum_{i=1}^5 \lambda_i r_i(y), \quad \text{with } r_5(y) = \texttt{STYLE-MST}(y).
\end{equation}
Overall, STYLE-MST acts as a memory-aware, population-level regularizer that allows the system to adaptively balance implicit privacy protection and stylistic expressiveness, ensuring that the generated text conforms to population-level stylistic norms while maintaining semantic coherence and quality.

\begin{algorithm}[t]
\caption{\textsc{Insert} Procedure for STYLE-MST}
\label{alg:stylemst}
\begin{algorithmic}[1]
\REQUIRE Sentence $s$; MST $\mathcal{T} = (V, E)$; metadata: embeddings \texttt{emb}, weights \texttt{weight}, parents \texttt{parent}, and minimum distances \texttt{minDist}
\ENSURE Updated MST $\mathcal{T}$
\STATE Compute embedding $\mathbf{e}_s \leftarrow \texttt{Embed}(s)$
\IF{$V = \emptyset$}
    \STATE Add $s$ to $V$ as root node
    \STATE Initialize \texttt{emb}, \texttt{weight}, \texttt{parent}, \texttt{minDist}
\ELSE
    \STATE Compute similarities $\mathrm{sim}(\mathbf{e}_s, \mathbf{e}_v)$ for all $v \in V$
    \STATE $i^\star \leftarrow \arg\max_v \mathrm{sim}(\mathbf{e}_s, \mathbf{e}_v)$
    \IF{$1 - \mathrm{sim}(\mathbf{e}_s, \mathbf{e}_{i^\star}) > \texttt{minDist}[i^\star]$}
        \STATE Add $s$ to $V$, connect to $i^\star$
        \STATE Set metadata: $\texttt{parent}[s] = i^\star$, $\texttt{weight}[s] = 1$
        \STATE Update $\texttt{minDist}[s]$ and all $\texttt{minDist}[v], v \in V \setminus \{s\}$
    \ELSE
        \STATE Merge $s$ into $i^\star$: $\texttt{weight}[i^\star] \mathrel{+}= 1$
        \STATE Update all $\texttt{minDist}[v], v \in V \setminus \{i^\star\}$
    \ENDIF
\ENDIF
\end{algorithmic}
\end{algorithm}

\section{Reinforcement Learning Training Loop} \label{sec:RL_trainingloop}

\subsection{Training Process}

We adopt a reinforcement learning framework based on GRPO to fine-tune a parameter-efficient LLaMA-3.2B model using the proposed composite reward. The model is initialized with 4-bit quantization and LoRA adapters targeting key transformer submodules (e.g., \texttt{q\_proj}, \texttt{k\_proj}, \texttt{o\_proj}, etc.), enabling memory-efficient training over long contexts.

During each iteration, the model generates $N$ candidate rewrites per input. These candidates are evaluated by the composite reward function described in Section~\ref{sec:reward}. The top-performing candidate is selected for reinforcement via a policy gradient update.

We configure the GRPO trainer with the following settings:
\begin{itemize}
    \item \textbf{Batching \& Sampling:} One example per device with four sampled completions per prompt; updates are aggregated via gradient accumulation.
    \item \textbf{Optimization:} AdamW optimizer with cosine learning rate schedule; warm-up ratio of 0.1 and weight decay of 0.1.
    \item \textbf{Model Adaptation:} LoRA rank of 16 with gradient checkpointing to enable efficient adaptation over long prompts.
    \item \textbf{Privacy Memory Update:} The most stylistically diverse and compliant candidate is added to the global memory structure (\texttt{STYLE-MST}) for future diversity scoring and outlier suppression.
\end{itemize}

We train the model for one epoch over the entire dataset, using a tunable mixture of reward components to supervise learning. During training, reward values and loss signals are logged in real-time via Weights \& Biases to monitor convergence and privacy-performance tradeoffs.

After training, the LoRA-adapted model and tokenizer are exported for downstream evaluation and deployment.


\subsection{Reward Function Design}\label{sec:reward}

To guide the learning of privacy-preserving text rewrites, we design a composite reward function that balances utility, privacy, and stylistic diversity. In contrast to prior methods that focus narrowly on semantic similarity or PII suppression, our formulation incorporates five orthogonal reward components that collectively capture both explicit and implicit privacy risks while ensuring textual fluency and format adherence.

     \textbf{Semantic Fidelity (Utility Guidance):}  
    We use BERTScore to quantify the semantic alignment between each rewritten candidate and the original reference text. This ensures that core meaning is preserved despite surface-level transformations. The BERTScore reward is normalized to lie in $[0, 1]$.

    \textbf{Length Consistency (Regularization):}  
    A token-based reward penalizes outputs that deviate significantly in length from the source. This is implemented as a Gaussian-shaped penalty:
    \begin{equation}
    r_{\text{length}}(y) = \exp\left( -\alpha \cdot \left( \frac{|y| - |x|}{|x|} \right)^2 \right)
    \end{equation}
    where $\alpha$ is a sharpness parameter and $|x|, |y|$ are input/output token lengths.

    \textbf{Format Compliance (Structure Guidance):}  
    Format adherence is checked via regex patterns or a finite-state checker that enforces structural constraints. Outputs that pass the check are rewarded with $+1$, and failures incur a penalty.

   \textbf{Entity Value Matching (Explicit Privacy Signal):}  
    Named Entity Recognition (NER) is applied to both input $x$ and output $y$ to identify explicit privacy signals, such as a person, a location, an organization, a product, etc. When there is an overlap, indicating memorization of the training data by the LLMs, a penalty is applied:
    \begin{equation}
    r_{\text{entity}}(y) = - \frac{|\text{NER}(x) \cap \text{NER}(y)|}{|\text{NER}(x)| + \epsilon}
    \end{equation}
    where $\epsilon$ avoids division by zero and scales the penalty to be robust for sparse entities.

\textbf{Style-Aware Privacy-Diversity Tradeoff (Implicit Signal):}  
We refine the STYLE-MST-based reward $r_{\text{MST}}(y)$ to reflect an adaptive balance between stylistic similarity and diversity based on whether the candidate is detected as a stylistic outlier. Let $\text{avg\_similarity}$ and $\text{close\_similarity}$ denote proximity to global and local memory nodes, and let $\text{diverse\_similarity}$ and $\text{bleu\_diverse\_similarity}$ capture stylistic and lexical novelty. The final reward adjusts based on the outlier status:

\begin{equation}
r_{\text{MST}}(y) = 
\begin{cases}
(s_{\text{avg}} + s_{\text{close}}) -  (d_{\text{sty}} + d_{\text{bleu}}), & \text{if } y \in \mathcal{O} \\
 (1 - s_{\text{avg}} -  (d_{\text{sty}} + d_{\text{bleu}}), & \text{otherwise}
\end{cases}
\end{equation}
where $s_{\text{avg}}$ and $s_{\text{close}}$ measure stylistic similarity, $d_{\text{sty}}$ and $d_{\text{bleu}}$ denote diversity penalties, and $\mathcal{O}$ is the set of outlier candidates.

\textbf{Formal Composition.}
Let $r_1$ through $r_5$ denote the five reward components described above. The final composite reward $R(y)$ for a candidate output $y$ is computed as:

\begin{equation}
R(y) = \sum_{i=1}^{5} \lambda_i \cdot r_i(y) \quad \text{where} \quad \sum_{i=1}^{5} \lambda_i = 1
\end{equation}

This composite reward is used as the training signal in the GRPO loop:
\begin{equation}
J(\theta) = \mathbb{E}_{y \sim \pi_\theta(\cdot \mid \mathbf{x}, \mathbf{c})}[R(y)]
\end{equation}

During training, for each prompt $x$, a batch of candidate completions $y_1, \dots, y_k$ is generated, scored individually via $R(y)$, and used to compute policy updates that optimize generation under both privacy and utility constraints.

This modular reward design supports dynamic reweighting, user personalization, and ablation-based attribution of model behavior. It further allows for plug-and-play addition of new privacy or utility signals without retraining the full model from scratch.

\subsection{Policy Optimization}
We formulate the privacy-preserving text rewriting task as a stochastic policy optimization problem, addressed through RL. Rather than selecting a single deterministic rewrite, the proposed approach learns a distribution over possible rewrites and optimizes this distribution to favor candidates that maximize privacy while preserving semantic meaning. Given an input prompt $\mathbf{x}$, let $\{y_1, y_2, \ldots, y_N\}$ denote a set of $N$ candidate rewrites generated by the model. Each candidate $y_i$ is evaluated by a reward function $R(y_i)$, which quantifies desirable properties such as reduced residual PII, implicit privacy information, diversity and semantic fidelity.

We define a parametric policy $\pi_\theta(y \mid \mathbf{x}, \mathbf{c})$ that models the probability of generating a rewritten text $y$ conditioned on an input prompt $\mathbf{x}$ and additional context information $\mathbf{c}$. The objective is to adjust the policy parameters $\theta$ to maximize the expected reward associated with generated rewrites. Formally, the objective function is given by:
\begin{equation}
J(\theta) = \mathbb{E}_{y \sim \pi_\theta(\cdot \mid \mathbf{x}, \mathbf{c})}[R(y)]
\end{equation}
where $R(y)$ denotes the reward assigned to each candidate rewrite, and the expectation is taken over the policy's output distribution.

Since $R(y)$ is typically non-differentiable with respect to $\theta$, we apply the REINFORCE algorithm to estimate the gradient of the objective. Using the log-likelihood trick, the gradient of $J(\theta)$ can be expressed as:
\begin{equation}
\nabla_\theta J(\theta) = \mathbb{E}_{y \sim \pi_\theta(\cdot \mid \mathbf{x}, \mathbf{c})} \left[ R(y) \nabla_\theta \log \pi_\theta(y \mid \mathbf{x}, \mathbf{c}) \right].
\end{equation}

In practice, this expectation is approximated by sampling $N$ candidate rewrites $y_i \sim \pi_\theta(\cdot \mid \mathbf{x}, \mathbf{c})$ and computing the empirical average:
\begin{equation}
\frac{1}{N} \sum_{i=1}^{N} R(y_i) \nabla_\theta \log \pi_\theta(y_i \mid \mathbf{x}, \mathbf{c}).
\end{equation}

This learning strategy gradually shifts the policy to assign higher probability to candidates that better satisfy privacy and semantic utility constraints.

This RL-based formulation offers two advantages over greedy selection strategies. First, it enables generalization across prompts: the learned policy captures privacy-preserving principles that apply broadly, reducing the need to generate and score multiple candidates per input. Second, it improves efficiency and scalability by eliminating costly inference-time search. Instead, the policy directly maps input text to optimized rewrites in a single pass, ensuring consistent application of privacy and utility criteria.

\subsubsection*{Policy Gradient Derivation}

To derive the gradient of the expected reward, we start with the objective function:
\begin{equation}
J(\theta) = \mathbb{E}_{y \sim \pi_\theta(\cdot \mid \mathbf{x}, \mathbf{c})}[R(y)].
\end{equation}

The REINFORCE algorithm uses the log-derivative trick, which allows us to move the gradient inside the expectation:
\begin{equation}
\nabla_\theta J(\theta) 
= \nabla_\theta \mathbb{E}_{y}[R(y)]
= \nabla_\theta \int \pi_\theta(y \mid \mathbf{x}, \mathbf{c}) R(y) \, dy.
\end{equation}

Applying the identity $\nabla_\theta \pi_\theta(y) = \pi_\theta(y) \nabla_\theta \log \pi_\theta(y)$, we get:
\begin{align}
\nabla_\theta J(\theta) 
&= \int \nabla_\theta \pi_\theta(y \mid \mathbf{x}, \mathbf{c}) R(y) \, dy \\
&= \int \pi_\theta(y \mid \mathbf{x}, \mathbf{c}) \nabla_\theta \log \pi_\theta(y \mid \mathbf{x}, \mathbf{c}) R(y) \, dy \\
&= \mathbb{E}_{y \sim \pi_\theta} \left[ R(y) \nabla_\theta \log \pi_\theta(y \mid \mathbf{x}, \mathbf{c}) \right].
\end{align}

This expectation is approximated via Monte Carlo sampling in practice, yielding the empirical gradient estimator:
\begin{equation}
\frac{1}{N} \sum_{i=1}^{N} R(y_i) \nabla_\theta \log \pi_\theta(y_i \mid \mathbf{x}, \mathbf{c}).
\end{equation}

\subsubsection*{DPO}
We adopt DPO, a contrastive fine-tuning method that aligns the model with human preferences without relying on a separate reward model. DPO directly optimizes a preference objective using pairwise comparisons between preferred and dispreferred responses, offering a simpler and fully differentiable alternative to RLHF.

DPO is a contrastive fine-tuning method designed for aligning language models with human preferences. Unlike reinforcement learning-based methods such as Reinforcement Learning from Human Feedback (RLHF), DPO directly optimizes a preference objective without relying on a separate reward model or policy sampling.

Given a dataset of triplets $(x, y_{\text{chosen}}, y_{\text{rejected}})$—where $x$ is the input, and $y_{\text{chosen}}$ and $y_{\text{rejected}}$ are model-generated responses annotated with relative preference—DPO fine-tunes the model to increase the likelihood of the preferred output while decreasing that of the dispreferred one. This is achieved by minimizing a KL-regularized contrastive loss between the fine-tuned model and a fixed reference model:
\[
\mathcal{L}_{\text{DPO}} = \log \left( \frac{\exp(\beta \cdot \Delta \log p)}{1 + \exp(\beta \cdot \Delta \log p)} \right),
\]
where $\Delta \log p = \log p_\theta(y_{\text{chosen}} \mid x) - \log p_\theta(y_{\text{rejected}} \mid x)$ and $\beta$ controls the sharpness of the contrast. The use of a reference model stabilizes training and prevents reward hacking.

DPO has emerged as a strong alternative to RLHF, offering comparable alignment quality with a simpler, fully differentiable optimization pipeline.

\subsection{Reasoning Process Integration}
To enhance transparency and training stability, the system incorporates an explicit reasoning step within its generative prompts. For every rewriting task, the model is prompted to generate a ``chain-of-thought" segment, typically enclosed by special tokens (e.g., \verb|<reasoning_start>| \ldots \verb|<reasoning_end>|).

This reasoning segment is structured to include:
\begin{enumerate}
    \item[(1)] An enumeration of all PII detected in the input text,
    \item[(2)] A list of any author attributes that can be inferred from the text, even if not explicitly stated (addressing implicit privacy), and
    \item[(3)] A description of the rewriting operations the model plans to apply to the text to enhance privacy or meet other objectives.
\end{enumerate}
Although this detailed reasoning is subsequently stripped from the final output made available to users or public platforms, its inclusion during the Reinforcement Learning (RL) training phase has been found to empirically stabilize policy updates. Furthermore, it significantly improves the interpretability of the model's behavior, offering insights into how it identifies privacy risks and decides on mitigation strategies.

\begin{table*}[t]
\centering
\resizebox{\textwidth}{!}{%
\begin{tabular}{l c | cccc | cccc | cccc}
\toprule
\textbf{Metric} &
\textbf{Goal} &
\multicolumn{4}{c|}{\textbf{Yelp}} &
\multicolumn{4}{c|}{\textbf{Twitter}} &
\multicolumn{4}{c}{\textbf{IMDb}} \\
\cmidrule(lr){3-6} \cmidrule(lr){7-10} \cmidrule(l){11-14}
& & Stylemix & DIPPER & DP-MLM & Ours & Stylemix & DIPPER & DP-MLM & Ours & Stylemix & DIPPER & DP-MLM & Ours \\
\midrule

\multicolumn{14}{l}{\textbf{Privacy}} \\
Entity~Match   & $\downarrow$ 
& 0.8038 & 0.3171 & \underline{0.2056} & \textbf{0.0809}
& 0.8179 & 0.2859 & \underline{0.1687} & \textbf{0.0784}
& 0.8221 & 0.3027 & \underline{0.1921} & \textbf{0.0835} \\

PEI            & $\downarrow$
& 0.0080 & 0.0060 & \underline{0.0020} & \textbf{0.0000}
& 0.0100 & \underline{0.0017} & 0.0020  & \textbf{0.0010}
& 0.0025 & 0.0048 & \underline{0.0020} & \textbf{0.0009} \\

Author~ID~F1   & $\downarrow$
& 0.6837 & 0.6737 & \underline{0.6640} & \textbf{0.5051}
& --     & --     & --     & --
& 0.7025 & 0.6812 & \underline{0.6695} & \textbf{0.4890} \\

Gender~F1      & $\downarrow$
& --     & --     & --     & --
& \underline{0.4970} & 0.5274 & 0.6027 & \textbf{0.3571}
& -- & -- & -- & -- \\

Outlier~Sim.   & $\downarrow$
& 0.0080 & \underline{0.0020} & 0.0060 & \textbf{0.0020}
& 0.0100 & \underline{0.0017} & 0.0020 & \textbf{0.0010}
& 0.0085 & \underline{0.0028} & 0.0056 & \textbf{0.0017} \\

\midrule
\multicolumn{14}{l}{\textbf{Meaning Preservation}} \\
BERTScore      & $\uparrow$
& \textbf{0.9303} & \underline{0.9196} & 0.8321 & 0.9164
& \textbf{0.9167} & 0.8878 & 0.8686 & \underline{0.8942}
& \textbf{0.9380} & \underline{0.9242} & 0.8465 & 0.9211 \\

MAUVE          & $\uparrow$
& \textbf{0.3754} & 0.0336 & 0.0221 & \underline{0.2522}
& \textbf{0.3450} & \underline{0.1287} & 0.1222 & 0.1276
& \textbf{0.3911} & 0.0452 & 0.0277 & \underline{0.2783} \\

\midrule
\multicolumn{14}{l}{\textbf{Diversity}} \\
Self\mbox{-}BLEU & $\uparrow$
& 0.7307 & 0.7602 & \textbf{0.8840} & \underline{0.7783}
& 0.6436 & \textbf{0.8026} & \underline{0.7956} & 0.7770
& 0.7205 & \textbf{0.8450} & \underline{0.8284} & 0.7902 \\

Lexical~Div.   & $\uparrow$
& 0.3958 & 0.6493 & \textbf{0.7728} & \underline{0.6687}
& 0.3294 & \underline{0.8097} & \textbf{0.8894} & 0.7824
& 0.4020 & \underline{0.7431} & \textbf{0.8090} & 0.7368 \\

\midrule
\multicolumn{14}{l}{\textbf{Utility}} \\
Sentiment~F1   & $\uparrow$
& \textbf{0.7792} & 0.7227 & 0.7231 &   \underline{0.7642}
& \underline{0.8130} & 0.6962 & 0.7500  & \textbf{0.8164}
& \textbf{0.8020} & 0.7614 & 0.7583  & \underline{0.7992} \\
\bottomrule
\end{tabular}
}
\caption{Evaluation results on Yelp, Twitter, and IMDb. Our method improves privacy metrics while maintaining strong semantic and utility performance. Arrows indicate desired directions. Best results are in \textbf{bold}; second-best are \underline{underlined}.}
\label{tab:main_results}
\end{table*}

\section{Experiment Setup} \label{sec:exp}
We evaluate our system on real-world datasets to measure performance along four core axes: privacy protection, meaning preservation, diversity and semantic fidelity.

\subsection{Datasets}
\label{sec:data}

\begin{table*}[t]
\centering
\small
\begin{tabular}{lcccccc}
\toprule
\textbf{Dataset} & \textbf{\#Train} & \textbf{\#Test} & \textbf{Avg.~Tokens} & \textbf{\% PII\,a} & \multicolumn{2}{c}{\textbf{Evaluation Tasks}} \\
\cmidrule(lr){6-7}
& & & & & \textbf{Privacy} & \textbf{Utility} \\
\midrule
Yelp    & 14{,}000 & 3{,}336 & 115.6 & 27.3 & Author ID (10-class) & Sentiment (2-class) \\
Twitter & 18{,}000 & 2{,}000 & 23.8  & 32.1 & Gender (2-class)     & Sentiment (2-class) \\
IMDb    & 12{,}000 & 1{,}500 & 220.7 & 34.8 & Author ID (10-class) & Sentiment (2-class) \\
\bottomrule

\end{tabular}
\caption{Dataset statistics and evaluation tasks. All datasets are used for binary sentiment classification (utility), and either author or gender classification (privacy).}
\label{tab:data_stats}
\end{table*}

We evaluated our privacy-preserving rewriting framework on three publicly available datasets that vary in domain, writing style, and implicit privacy risks. Each data set is used to benchmark two tasks: binary sentiment classification and author-related prediction (author ID or gender classification), enabling a comprehensive assessment of semantic utility and privacy preservation.

 \textbf{Twitter.} We use a subset of the Twitter dataset originally compiled for training a CrowdFlower gender classifier. Annotators were asked to label users based on their Twitter profile as either male, female, or brand. We extract tweets from users with binary gender labels (male or female) and discard brand accounts. The dataset contains 20,000 entries, from which we retain 18,000 for training and 2,000 for testing. Sentiment labels are assigned using a pretrained classifier. The text is short-form and informal, with an average of 24 tokens per tweet.

\textbf{IMDb.} The IMDb dataset consists of 62,000 user reviews by 62 authors on IMDb. Following prior work~\cite{utpala2023locallydifferentiallyprivatedocument}, we use a 10-author subset with 13,500 reviews total (1,350 per author). These long-form movie reviews are subjective and stylistically rich. We assign binary sentiment labels using a sentiment classifier and use author ID for the 10-class prediction task.

\textbf{Yelp.} The Yelp dataset \cite{yelp2017} includes 17,336 customer reviews across restaurants and businesses. From the full set, we sample 10 authors, each contributing roughly 1,700 reviews. Reviews exceeding 512 tokens are removed for compatibility with LLM input length. Sentiment labels are generated automatically and manually verified for class balance. The average length is 116 tokens per review.

Each dataset is evaluated on two supervised tasks: binary sentiment classification and a secondary privacy-sensitive classification, 10-way author identification for IMDb and Yelp, and 2-way gender classification for Twitter.

\subsection{Experimental Setup}
\label{sec:setup}

\textbf{Model and Training.}
We initialize our rewriting model from \texttt{Llama-3.2-3B-Instruct-bnb-4bit}, a 3.2-billion parameter variant of LLaMA-3 quantized with 4-bit GPTQ for efficient memory usage. A 32-rank LoRA adapter is integrated and fine-tuned using GRPO, a reinforcement learning framework that jointly optimizes privacy, semantic fidelity, and downstream utility.

Training is conducted with a learning rate of $5 \times 10^{-5}$, 16 gradient accumulation steps, and 10 RL epochs. For each input prompt, 16 candidate generations are sampled per iteration. All experiments are run on a single NVIDIA A100 (80GB) GPU. Flash-attention and quantized weights limit peak memory usage to under 70\%. All random seeds are fixed to 3407 to ensure reproducibility. Further training details can be found in Appendix~\ref{app:hyper}.

\textbf{Baselines.}
We compare our method against three recent state-of-the-art privacy-preserving text rewriting models:
\begin{itemize}
    \item \textbf{Stylemix(2024)}~\cite{fisher-etal-2024-styleremix}: a style manipulation model that alters sentence structure and lexical choice to obfuscate author identity while retaining semantic meaning.
    \item \textbf{DIPPER(2024)}~\cite{krishna2024paraphrasing}: a controllable paraphrasing framework that disentangles privacy and utility dimensions to support adjustable rewriting.
    \item \textbf{DP-MLM(2024)}~\cite{meisenbacher2024dpmlmdifferentiallyprivatetext}: a differentially private masked language model designed to provide strong theoretical privacy guarantees during generation.
\end{itemize}

\textbf{Data Preprocessing.}
All input texts are tokenized using the default \texttt{spaCy} English tokenizer. Samples exceeding 512 tokens are discarded. To ensure class balance in downstream evaluations, we uniformly downsample data across sentiment and author ID labels. Each data set is divided into subsets 80\% training, 10\% validation, and 10\% testing. Personally Identifiable Information (PII) spans are detected using spaCy NER augmented with rule-based entity matching patterns.

\textbf{Downstream Evaluation and Privacy Utility}
We assess the effectiveness of rewriting using two downstream classification tasks:
\begin{itemize}
    \item \textbf{Sentiment Classification} (Yelp, Twitter): Classifies text sentiment as positive, neutral, or negative.
    \item \textbf{Author Identification} (Yelp, IMDb): Serves as a proxy for privacy leakage: Higher prediction accuracy implies greater retention of author-specific characteristics.
\end{itemize}

For each task, we fine-tune a \texttt{bert-base-cased} classifier using the HuggingFace \texttt{Trainer}. The configuration includes a batch size of 16, 5 training epochs, and early stopping based on macro-F1. Evaluation metrics include precision, F1 score, and Matthews correlation coefficient.

Input sequences are tokenized with padding and truncation to a fixed maximum length. We train individual classifiers for each label field (sentiment or author ID) using the rewritten text and report results on original samples to measure semantic preservation and anonymization effectiveness.

\begin{figure}[t]
  \centering
  \includegraphics[width=0.5\textwidth]{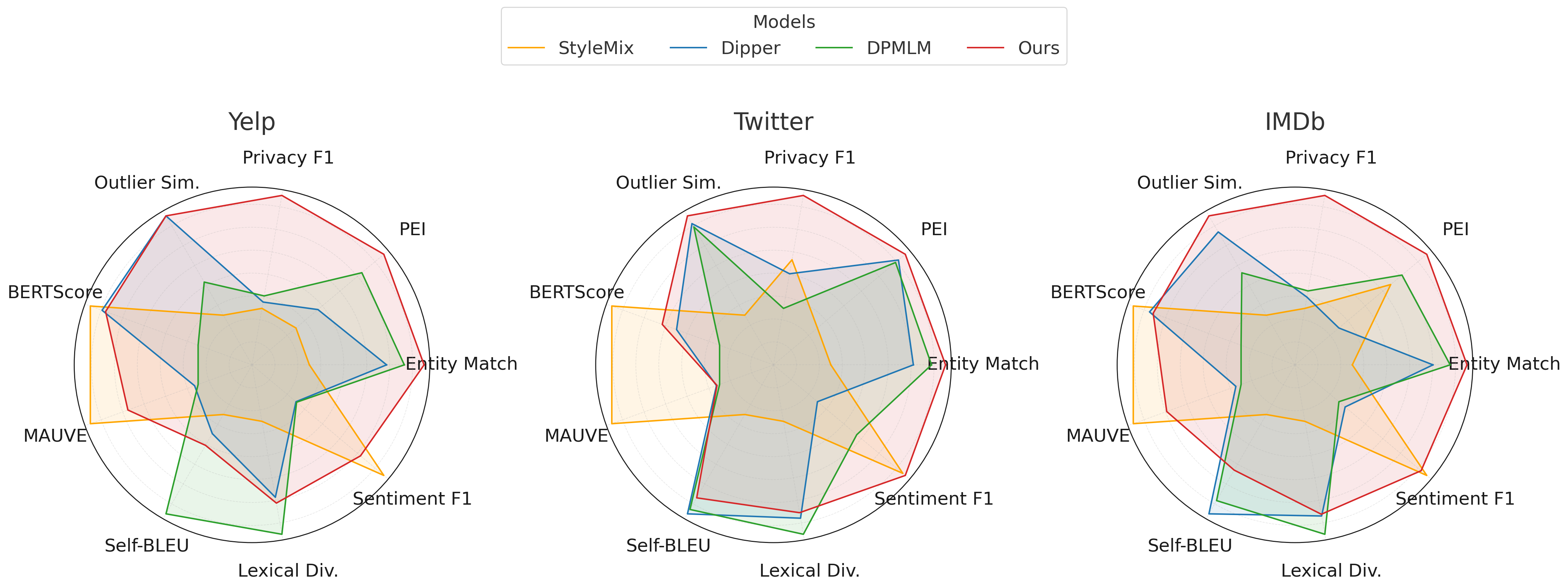}
  \caption{
    \textbf{Radar plot comparison across Yelp, Twitter, and IMDb.}
    Each axis represents a normalized metric (higher is better), covering privacy (Privacy~F1), semantic fidelity (BERTScore), diversity (Self-BLEU and Lexical Diversity), and utility (Sentiment~F1). Our method consistently achieves the most balanced performance across datasets, with leading or second-best results across all metrics. For privacy, we use Author~F1 when available (Yelp, IMDb) and Gender~F1 otherwise (Twitter).
    }
    \label{fig:radar_plot}
\end{figure}

\subsection{Evaluation Metrics}
\label{sec:metrics}
To conduct a holistic evaluation, we employ a suite of complementary metrics designed to quantify the performance of our system along three critical axes: semantic fidelity, privacy protection, and linguistic quality.

\textbf{\textsc{BertScore} ($\uparrow$)}
This metric measures the semantic similarity between a reference text $x$ and a candidate rewrite $y$ by computing the cosine similarity of their token embeddings from a pretrained model like RoBERTa. We report the F1 measure, defined as:
\begin{equation}
    \text{BERTScore} = 2 \frac{P_{\text{BERT}} \cdot R_{\text{BERT}}}{P_{\text{BERT}} + R_{\text{BERT}}},
\end{equation}
where $P_{\text{BERT}}$ and $R_{\text{BERT}}$ are the precision and recall calculated over token-level similarities. Higher scores indicate better semantic preservation.

\textbf{\textsc{Mauve} \cite{darling2004mauve} ($\uparrow$)}
This metric quantifies the divergence between the distribution of generated texts and reference texts. It is estimated via the KL divergence between these distributions in a deep latent space. Higher values imply that the model's output distribution more closely aligns with human-written text.

\textbf{\textsc{Sentiment F1} ($\leftrightarrow$)}
We use the Macro-F1 score of a pretrained sentiment classifier (e.g., \texttt{cardiffnlp/twitter-roberta-base-sentiment}) on the rewritten text to proxy for downstream utility. The goal is to maintain the original sentiment, so a score close to the original text's performance is ideal.

\textbf{\textsc{Entity Match} ($\downarrow$)}
This metric calculates the recall of PII entities from the original text $x$ that persist in the rewritten text $y$. Letting $E(x)$ be the set of PII entities in the original, the metric is:
\begin{equation}
    \text{Entity Match} = \frac{|E(x) \cap E(y)|}{|E(x)|}.
\end{equation}
Lower values denote stronger de-identification and better privacy protection.

\textbf{\textsc{Privacy Exposure Index (PEI)} ($\downarrow$)}
PEI measures the empirical probability that a black-box ranker $\mathcal{R}$ can successfully identify the original text $x$ from a large corpus $\mathcal{C}$ given only the rewrite $y$. It is formally expressed as:
\begin{equation}
    \text{PEI} = P(\mathcal{R}(y, \mathcal{C}) = x).
\end{equation}
A lower PEI indicates stronger resistance to membership inference attacks.

\textbf{\textsc{Author-ID F1} ($\downarrow$)}
To measure stylometric privacy, we use the Macro-F1 score of a style-based author classifier evaluated on the rewritten texts. A significant drop in F1 compared to the performance on original texts suggests successful style obfuscation.

\textbf{\textsc{Outlier Similarity} ($\downarrow$)}
This metric quantifies the maximum cosine similarity between the style embedding of a candidate rewrite $\mathbf{v}_y$ and the set of embeddings from previously flagged stylistic outliers $\mathcal{O}$.
 \begin{equation}
    \text{Outlier Sim} = \max_{\mathbf{v}_o \in \mathcal{O}} \frac{\mathbf{v}_y \cdot \mathbf{v}_o}{\|\mathbf{v}_y\| \|\mathbf{v}_o\|}.
 \end{equation}
Lower scores indicate that the model successfully avoids the replication of risky outlier writing styles.

\textbf{\textsc{Self-BLEU} ($\downarrow$)}
This metric measures n-gram diversity by calculating the average pair-wise BLEU-4 score across the set of all generated sentences $S = \{y_1, \dots, y_N\}$. A lower score indicates less n-gram overlap and thus greater lexical diversity.
\begin{equation}
    \text{Self-BLEU} = \frac{1}{N(N-1)} \sum_{i=1}^N \sum_{j \neq i} \text{BLEU}(y_i, \{y_j\}).
\end{equation}

\textbf{\textsc{Lexical Diversity (TTR)} ($\uparrow$)}
The Type-Token Ratio is a straightforward measure of vocabulary richness, calculated as the number of unique words (types) divided by the total number of words (tokens) in the generated corpus.
\begin{equation}
    \text{TTR} = \frac{|\text{Unique Tokens}|}{|\text{Total Tokens}|}.
\end{equation}
Higher values reflect a richer and more varied vocabulary in the output.


\section{Results}\label{sec:result_analysis}

We evaluate our privacy-preserving rewriting framework against three strong baselines: Stylemix(2024), DIPPER(2024), and DP-MLM(2024) on the Yelp, Twitter, and IMDb datasets, using a comprehensive suite of metrics that assess semantic fidelity, diversity, privacy, and utility (Table~\ref{tab:main_results}).

\begin{figure}[t]
  \centering
  \includegraphics[width=0.5\textwidth]{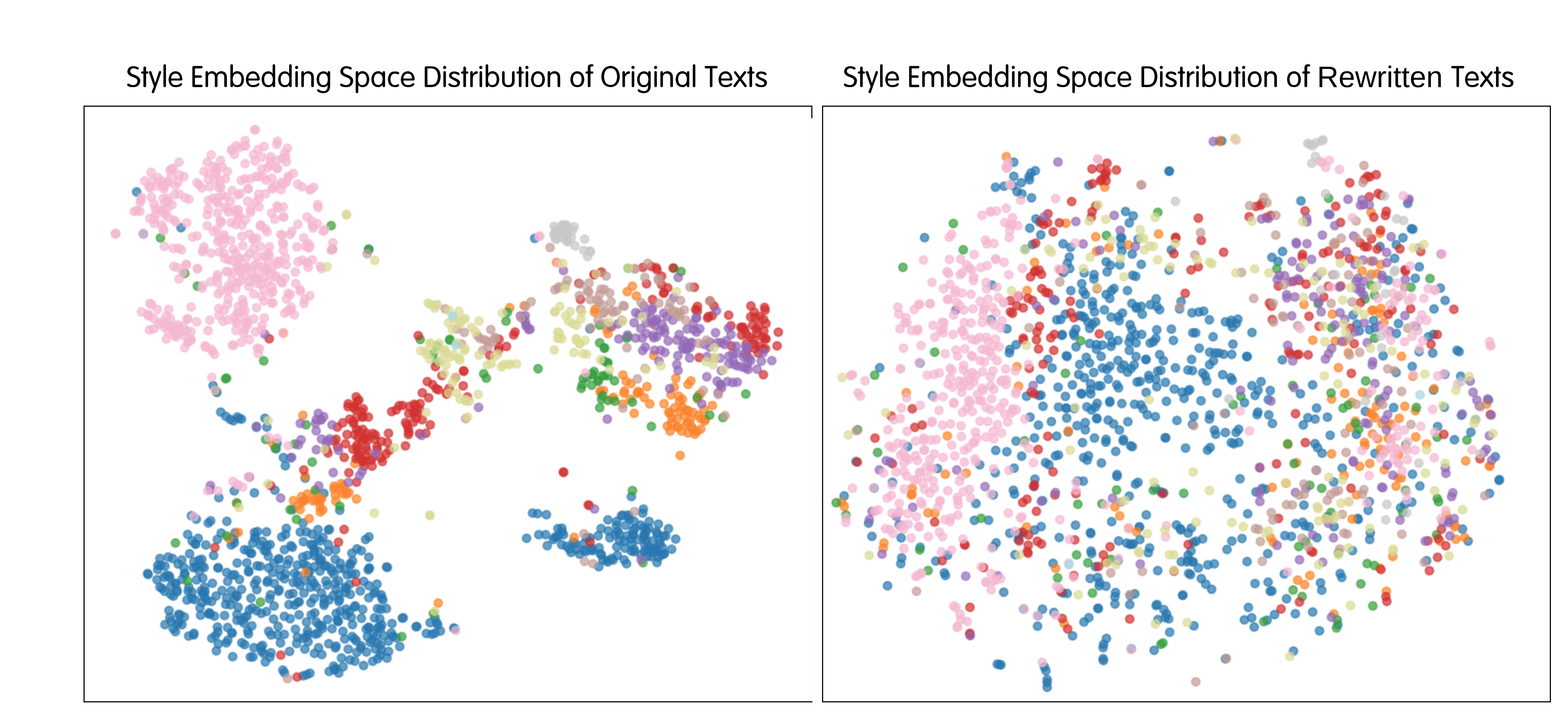}
  \caption{t-SNE visualization of style embeddings. Left: Original text shows tight author-based clusters. Right: Rewritten text shows dispersed patterns, indicating reduced identifiability and enhanced stylistic anonymization.}
  \label{fig:emebdding_dif}
\end{figure}

\subsection{Quantitative Performance Analysis}
\textbf{Privacy performance} is where our model shows the most significant gains. It achieves the lowest Entity Match scores across all datasets (e.g., 0.0809 on Yelp and 0.0784 on Twitter), indicating strong suppression of explicit PII. Additionally, our method reduces PEI to near-zero or zero values, and substantially lowers downstream reidentification metrics: author F1 drops to 0.5051 on Yelp and 0.4890 on IMDb, while gender F1 on Twitter is reduced to 0.3571, outperforming all baselines. These results demonstrate the ability of the framework to protect explicit and implicit identity traces.

In terms of \textbf{semantic fidelity}, our model achieved good performance on BERTScore and MAUVE across all datasets, trailing only Stylemix (e.g., 0.9164 vs 0.9303 BERTScore on Yelp). Despite this small gap, our model achieves MAUVE higher than DIPPER and DP-MLM, reflecting better distributional alignment between original and rewritten texts without relying on copying.

For \textbf{diversity}, our method consistently achieves second-best Self-BLEU and lexical diversity scores on Yelp, it reaches 0.7783 Self-BLEU and 0.6687 lexical diversity, slightly behind the noisier DP-MLM output. For Twitter and IMDb dataset, the performance difference between  our model and top performance are also marginal. Meanwhile, although DP-MLM shows greater diversity (e.g. Self-BLEU 0.8840), it suffers from reduced semantic alignment, highlighting our model's more balanced handling of style variation and coherence. The observed overdiversify in DP-MLM stems from token-level DP sampling, which introduces high variance in output structure.

Finally, the model maintains strong \textbf{utility}, with Sentiment~F1 scores competitive across all datasets. In particular, it achieves 0.7642 on Yelp (second-best), 0.8164 on Twitter (best), and 0.7992 on IMDb (second-best), confirming that task-relevant signals such as sentiment and emotional tone are preserved despite privacy transformations.

In sum, given the and Figure~\ref{fig:radar_plot}) the model maintains utility despite these privacy safeguards. The sentiment classification F1 scores on the rewritten texts remain comparable to or better than those from the baselines, showing that the user's emotional intent and task-relevant features are preserved. Collectively, these results validate that our reward-based learning framework is capable of maintaining a strong trade-off between privacy and usability, achieving best-in-class performance in several categories and maintaining second-best elsewhere without degradation in quality.

\begin{table}[h]
\centering
\footnotesize
\begin{tabular}{p{0.95\columnwidth}}
\toprule
\textbf{Original Text:} \\
\textit{This is definitely one of the most neighborhood of all neighborhood bars. It is just like a suburban house set amongst many other houses. You won't be getting all rowdy and crazy in here but you can certainly have a calm and collected good time. When people move to Old Metairie and miss their nasty days at Snake'n Jakes they can come here for a clean and hygienic version of a hidden away neighborhood bar.} \\
\midrule
\textbf{Detected PII:} Old Metairie (location), Jakes (name) \\
\textbf{Inferred Attributes:} Age (21–30), Gender (likely male), Cultural familiarity with bars \\
\midrule
\textbf{Rewritten Output:} \\
\textit{This is definitely one of the most discreetly located establishments. It is just like an unassuming residence set amongst many others. You won't be getting all boisterous and unruly in here but you can certainly have a composed and serene experience. When individuals relocate to the Oakwood district and yearn for a refined and sanitary atmosphere comparable to a secluded residential outpost, they can come here for a clean and private variant of a concealed residential gathering spot.} \\
\bottomrule
\end{tabular}
\caption{Example of privacy-preserving rewriting. Explicit PII and implicit demographic cues are removed or neutralized while preserving semantics.}
\label{tab:sample-examples}
\end{table}

\subsection{Qualitative Performance Analysis}
To complement the quantitative metrics, we provide a qualitative analysis of our rewriting framework by examining both individual rewriting cases and distributional shifts in the author style space. These examples help illuminate how our method achieves privacy preservation without compromising semantics or fluency.

\vspace{0.5em}
\noindent \textbf{Example Analysis.}
Table~\ref{tab:sample-examples} (Appendix~\ref{app:examples}) presents a representative rewriting instance. The original text contains both explicit identifiers (e.g., \textit{Old Metairie}, \textit{Jakes}) and implicit authorial cues, such as informal phrasing and references to a specific cultural context. In the rewritten output, these cues are systematically neutralized: location names are abstracted (\textit{Old Metairie} $\rightarrow$ \textit{Oakwood district}), idiomatic expressions are replaced with more neutral language (e.g., \textit{nasty days} $\rightarrow$ \textit{refined atmosphere}), and stylistic informality is reduced to convey a composed and generic tone.

This example illustrates how our model performs controlled abstraction and paraphrasing to achieve stylistic obfuscation. The semantic core (e.g., describing a low-key neighborhood bar) is preserved, while privacy-sensitive details and stylistic outliers are softened or restructured. More rewriting examples are included in Appendix~\ref{app:examples}.

\vspace{0.5em}
\noindent \textbf{Embedding Visualization.}
Figure~\ref{fig:emebdding_dif} presents 2D t-SNE projections of author embeddings before and after rewriting. In the left panel (original text), strong author-specific clusters emerge, highlighting the distinctiveness and traceability of writing styles in the raw data. This clustering indicates a high risk of re-identification through stylometry.

After rewriting (right panel), the distribution becomes significantly more diffused and intermixed. Clusters dissolve, suggesting reduced correlation between embedding positions and author identity. Crucially, this dispersion is not uniform noise: it is structured by the STYLE-MST module, which encourages alignment with population-level stylistic distributions while avoiding homogeneity.

When inputs are flagged as stylistic outliers, our model applies soft regularization toward stylistic centroids in the memory graph. When inputs lie near the stylistic mean, the model promotes diverse yet semantically grounded rewrites. This mechanism ensures that rewritten outputs remain expressive and task-relevant, while making authorship inference significantly more difficult.

\vspace{0.5em}
\noindent In sum, both the rewriting examples and embedding visualizations underscore our model's ability to achieve nuanced privacy transformations that go beyond simple token redaction, re-embedding the text into a safer, anonymized stylistic manifold.

\begin{table}[t]
\centering
\footnotesize
\setlength{\tabcolsep}{2.5pt} 

\begin{tabular}{lcccc}
\toprule
\textbf{Model} & \textbf{BERTScore}$\uparrow$ & \textbf{Privacy-F1}$\downarrow$ & \textbf{Self-BLEU}$\uparrow$ & \textbf{Sentiment-F1}$\uparrow$ \\
\midrule
Base           & 0.9021 & 0.7450 & 0.7010 & 0.7321 \\
+RL            & 0.9154 & 0.6222 & 0.7253 & 0.7550 \\
+Context       & 0.9187 & 0.6084 & 0.7406 & 0.7601 \\
+Memory        & 0.9164 & 0.5488 & 0.7783 & 0.7642 \\
+Reasoning     & 0.9164 & 0.5051 & 0.7783 & 0.7642 \\
\bottomrule
\end{tabular}
\caption{Ablation results of LLM-based rewriting. Each row adds a module incrementally. Lower Privacy-F1 indicates stronger anonymization.}
\label{tab:ablation_concise}
\end{table}

\subsection{Ablation Analysis}

Table~\ref{tab:ablation_concise} presents an incremental ablation study to evaluate how each component contributes to the performance of our rewriting framework on the Yelp dataset. Starting from the base language model (\textbf{Base}), we progressively add reinforcement learning (\textbf{+RL}), in-context prompting (\textbf{+Context}), memory-based style modeling (\textbf{+Memory}), and finally a reasoning-guided RL mechanism (\textbf{+Reasoning}).

The \textbf{Base} model, with no auxiliary guidance, exhibits weak privacy protection (Privacy-F1 = 0.7450) and limited diversity (Self-BLEU = 0.7010). Although it maintains decent semantic and sentiment performance, its outputs retain many author-identifiable traits.

Adding \textbf{RL} alone improves performance across all metrics, reducing Privacy-F1 to 0.6222 and increasing sentiment alignment. However, privacy remains suboptimal, indicating that the RL objective requires stronger contextual signals to achieve robust anonymization.

Introducing \textbf{Context} yields further gains. Privacy-F1 drops to 0.6084, and Self-BLEU rises to 0.7406, suggesting the model benefits from being guided by stylistic or structural cues embedded in the prompt.

Adding \textbf{Memory} introduces population-level style awareness, resulting in a substantial reduction in Privacy-F1 to 0.5488 and a notable boost in diversity (Self-BLEU = 0.7783). This confirms that memory helps the model adapt to global stylistic distributions and better obfuscate author-specific patterns.

Finally, integrating the \textbf{Reasoning} module enhances privacy even further (Privacy-F1 = 0.5051), while preserving semantic quality (BERTScore = 0.9164) and diversity. This configuration achieves the strongest trade-off across all evaluation dimensions, demonstrating that structured, reward-informed reasoning helps refine rewrites with respect to both privacy and utility.

Overall, these results highlight the importance of combining multilevel guidance, reinforcement, context, memory, and reasoning, to effectively balance semantic fidelity, privacy, diversity, and task relevance.

\section{Conclusion and Future Work} \label{sec:conclusion}

This paper presents a reinforcement learning framework for privacy-preserving text rewriting that addresses both explicit and implicit privacy threats. By integrating a style-aware outlier detector, a graph-regularized reward function, and a composite scoring scheme, the proposed approach enables fine-grained control over privacy–utility trade-offs in natural language generation. The proposed system leverages a LoRA-adapted LLaMA model fine-tuned via GRPO to generate semantically faithful rewrites that minimize exposure to PII and stylometric re-identification. Empirical evaluations demonstrate that the proposed method outperforms strong baselines in author anonymization, entity suppression, and implicit privacy metrics, while preserving task relevance and linguistic richness. This work highlights the importance of incorporating population-level stylistic regularization and structured reinforcement signals into text generation pipelines aimed at enhancing user privacy.

Several directions remain open for improving the robustness, adaptability, and generalization of the proposed framework. First, future efforts will focus on incorporating length-sensitive and context-aware normalization strategies to improve style and privacy scoring, particularly for shorter or highly variable inputs such as tweets. Second, the fixed reward weighting scheme used in this work could be extended to a dynamic, user-adaptive formulation that allows end-users or downstream systems to specify desired trade-offs between semantic fidelity and privacy risk. Another important avenue is extending the framework to support multilingual data, which will require adapting the style embeddings, entity detectors, and reward components to linguistically diverse and low-resource settings. In addition, while the proposed method shows strong empirical privacy performance, integrating formal differential privacy mechanisms into the optimization loop remains an important step toward offering provable guarantees. Finally, we plan to conduct human evaluations, including red-team audits and user studies, to assess real-world privacy perception and utility. These extensions will help advance the development of privacy-aware language generation systems that are both secure and deployable across diverse applications.

\bibliographystyle{IEEEtran}
\bibliography{references}
\clearpage

\appendices
\section{System Prompt for Privacy-Preserving Rewriting}
\label{appendix:system_prompt}

The system prompt used to guide the rewriting model is detailed below. It directs the model to detect both explicit and implicit personal identifiers, anonymize sensitive content, and ensure semantic consistency in the rewritten text.

\begin{quote}
\texttt{
You are a sophisticated privacy-focused text anonymizer. Task is to rewrite input text to protect personal information and maintain confidentiality. \\
Firstly, detect all personally identifiable information (PII) such as names, email addresses, phone numbers, physical addresses, and identification numbers. \\
And imply implicit author attribute like age, gender, and writing style; they can guide your rewrite. \\
Place your reasoning process between \{reasoning\_start\} and \{reasoning\_end\}. \\
Then, replace each identified piece of PII with a fake value that matches the type and context. \\
Please change the text writing style and hide all personal attributes. Remove any informal language, including slang, idioms, or personal tone. \\
The rewritten text needs to reserve meaning. You can adjust nominal words and phrases as necessary while preserving the core meaning of the original text. Ensure that the rewritten text conveys the same overall message and intent. \\
Place your rewritten text between \{solution\_start\} and \{solution\_end\} after the reasoning part.
}
\end{quote}

\section{Training Hyperparameters}
\label{app:hyper}

\subsection{GRPO Model Training Configuration}

We provide the complete hyperparameter configuration used for fine-tuning our LLM-based rewriter with GRPO (Generative Reinforcement-learning from Proxy Objectives). The model is optimized using the AdamW optimizer with fused kernels for performance efficiency.

\begin{itemize}
    \item \textbf{Learning rate:} 5e-5
    \item \textbf{Optimizer:} \texttt{adamw\_torch\_fused}
    \item \textbf{Adam $\beta_1$, $\beta_2$:} 0.9, 0.99
    \item \textbf{Weight decay:} 0.1
    \item \textbf{Warmup ratio:} 0.1
    \item \textbf{Scheduler:} Cosine decay
    \item \textbf{Batch size:} 1 per device, with 16-step gradient accumulation
    \item \textbf{Max prompt length:} 256 tokens
    \item \textbf{Max generation length:} $L - 256$ (where $L$ is the sequence max length)
    \item \textbf{Number of candidate generations:} 16 per prompt
    \item \textbf{Epochs:} 10 full RL epoch
    \item \textbf{Max gradient norm:} 0.1
    \item \textbf{Logging frequency:} Every step
    \item \textbf{Save frequency:} Every 100 steps
\end{itemize}
This configuration was selected to balance sample diversity and stability, while ensuring compatibility with single-GPU 4-bit quantized LLMs.

\subsection{Downstream Classifier Fine-Tuning}

For sentiment and author ID classification tasks, we fine-tune a \texttt{bert-base-cased} model using the HuggingFace \texttt{Trainer} API with the following configuration:

\begin{itemize}
    \item \textbf{Batch size:} 16 for both training and evaluation
    \item \textbf{Number of epochs:} 5
    \item \textbf{Evaluation strategy:} At the end of each epoch
    \item \textbf{Model saving strategy:} Save best-performing checkpoint based on macro-F1
    \item \textbf{Metric for best model:} F1 Score
    \item \textbf{Early stopping:} Enabled via best model tracking
    \item \textbf{Logging frequency:} Every 10 steps
    \item \textbf{Reporting:} Disabled for reproducibility (\texttt{report\_to=none})
\end{itemize}

The classifier models are trained separately for each evaluation task (sentiment, author ID, or gender prediction), and evaluated using accuracy, macro-F1, and Matthews correlation coefficient.
\clearpage

\section{Sample Rewriting Examples}
\label{app:examples}

To better illustrate our model’s privacy-preserving rewriting behavior, Table~\ref{tab:sample-examples} presents qualitative examples, including detected PII, inferred attributes, the model’s reasoning, and the final rewritten output.

\begin{table}[h]
\centering
\footnotesize
\begin{tabular}{p{0.95\columnwidth}}
\toprule
\textbf{Example 2} \\
\textbf{Original Text:} \\
\textit{My cousin Jason and I spent the summer of 2002 exploring back alleys in Brooklyn and drinking cold beers on rooftops. We even went to that illegal rave near Flatbush where everyone wore glow sticks.} \\
\textbf{Detected PII:} Jason (name), 2002 (date), Brooklyn, Flatbush (locations) \\
\textbf{Inferred Attributes:} Age (late 30s–40s), Cultural preferences (underground music), Male bonding \\
\textbf{Rewritten Output:} \\
\textit{A friend and I spent one summer in the early 2000s exploring hidden spots in a northeastern city and enjoying chilled beverages under the stars. One highlight included a late-night music event in an industrial district, filled with glowing lights and energetic dancing.} \\
\midrule

\textbf{Example 3} \\
\textbf{Original Text:} \\
\textit{Dr. Kim Nguyen, who taught my psychology class at Stanford in 2015, changed my perspective on cognitive biases. Her lectures were so good I still reference them in my work at the VA hospital.} \\
\textbf{Detected PII:} Kim Nguyen (name), Stanford (organization), 2015 (date), VA hospital (employer) \\
\textbf{Inferred Attributes:} Education (graduate level), Profession (healthcare or psychology), Age (30s–40s) \\
\textbf{Rewritten Output:} \\
\textit{A professor I studied under during my academic years deeply influenced my understanding of psychological phenomena. Their insights continue to inform my professional work in the healthcare sector.} \\
\midrule

\textbf{Example 4} \\
\textbf{Original Text:} \\
\textit{We always used to grab bubble tea from that corner shop on 12th Street in Austin before heading to open mic night at Paper Tiger. It was our Friday ritual back in college.} \\
\textbf{Detected PII:} 12th Street, Austin (location), Paper Tiger (venue), college timeline \\
\textbf{Inferred Attributes:} Age (20s), Education (student), Region (Southern U.S.) \\
\textbf{Rewritten Output:} \\
\textit{We often visited a small drink shop on a busy street before attending live performances at a local venue. This weekly routine was a memorable part of our younger academic years.} \\
\bottomrule
\end{tabular}
\caption{Illustrative examples showing detected PII, inferred attributes, and rewritten outputs that preserve utility while obfuscating sensitive information.}
\label{tab:rewriting-examples}
\end{table}

\end{document}